# S&P500 FORECASTING AND TRADING USING CONVOLUTION ANALYSIS OF MAJOR ASSET CLASSES

Panagiotis Papaioannou[1], Thomas Dionysopoulos[2], Dietmar Janetzko[3], Constantinos Siettos[1,*]

[1]School of Applied Mathematics and Physical Sciences, National Technical University of Athens, Greece

*Corresponding author

[2]AXIANTA RESEARCH, Nicosia, Cyprus - Avenir Finance Investment Managers, Paris, France

[3]Cologne Business School, Cologne, Germany

## Abstract

By monitoring the time evolution of the most liquid Futures contracts traded globally as acquired using the Bloomberg API from 03 January 2000 until 15 December 2014 we were able to forecast the S&P 500 index beating the "Buy and Hold" trading strategy. Our approach is based on convolution computations of 42 of the most liquid Futures contracts of four basic financial asset classes, namely, equities, bonds, commodities and foreign exchange. These key assets were selected on the basis of the global GDP ranking across countries worldwide according to the lists published by the International Monetary Fund (IMF, Report for Selected Country Groups and Subjects, 2015). The main hypothesis is that the shifts between the asset classes are smooth and are shaped by slow dynamics as trading decisions are shaped by several constraints associated with the portfolios allocation, as well as rules restrictions imposed by state financial authorities. This hypothesis is grounded on recent research based on the added value generated by diversification targets of market participants specialized on active asset management, who try to efficiently and "smoothly navigate" the market's volatility.

# 1. INTRODUCTION

Corporate finance theorists, macroeconomists, behavioral psychologists, quantitative finance mathematicians have teamed up to find the holy grail of the financial markets: that of consistently beating the market. Towards this aim, methods can be categorized into (a) fundamentals, and (b) financial engineering. On the one hand, Fundamentals include models from corporate finance theories, (e.g. Capital Asset Pricing Model: Fama and French, 1989) and macroeconomic theories (e.g. the Keynesian model: Keynes, 1936; Bodkin and Eckstein, 1985 etc.) most of which try to approximate the dynamics of macroeconomic observables such as the aggregate demand and supply, investment volume and consumption, risk premia etc. On the other hand, financial engineering and quantitative finance models, that have been flourished after the seminal work of the derivatives contracts pricing model (Black and Scholes, 1973) try to forecast the price action of several financial instruments such as the S&P 500 Index (Niaki and Hoseinzade, 2013) on a data driven basis.

Both approaches focus basically on several sub-targets such as the forecasting of a single asset as well as asset allocation methods among the four basic asset classes, namely, Equities, Commodities, Bonds, and Foreign Exchange Markets (Bekkers, et al., 2012). Several models have been proposed to forecast their future dynamics based on historical data and information acquired from almost all possible sources: price action, fundamentals as well as behavioral sentiment analysis and e-social platforms (Husain and Bowman, 2004; Polk, et al., 2006; Diebold and Li, 2006; Papaioannou et al., 2013). Even though success stories have been reported, these models are still prone to failure in situations undergoing structural changes and market crashes

(Zellner&Chetty, 1965; Klein &Bawa, 1976; Kandel&Stambaugh, 1996; Barberis, 2000; Caballero et al., 2008).

Here we propose an approach to forecast the E-mini S&P500 Futures contract exploiting convolution analysis of 42 of the most liquid Futures contracts of four basic financial asset classes: (1) equities, (2) bonds, (3) commodities, and, (4) foreign exchange.Our choice is motivated by the fact that we want to use a "non-memory" approach, one that does not need training datasets (e.g. ANNs etc.) and which would provide a pointwise, time dependent and out-of-sample trading strategy even from the first few observations of the historical dataset. Based on the proposed approach, we managed to successfully forecast the S&P500 Futures Contract beating the "Buy and Hold" benchmark trading strategy.

## 2. METHODOLOGY

### 2.1 The Hypothesis

The question of how financial uncertainty gets incorporated in the risk premia offered by several financial assets and how it formulates the investors' preferences and their corresponding portfolios allocation is fundamental in contemporary financial research. Previous studies have shown that allocation decisions made by fund managers on behalf of the investors are shaped by several bounds regarding the portfolios weights of each asset class in order to ensure portfolio robustness, lower transaction costs (turnovers), longer investment horizon, smaller concentrations etc. (Roncalli, 2010; Roncalli and Weisang, 2015; Jagannathan and Ma, 2003).

Building up on these studies, our main hypothesis is that the shifts between the combinations of asset classes must be smooth and relatively slow, in order to achieve solid, safe and robust allocation of capital across economies and markets. An argument that supports this hypothesis is that the investors rebalance their portfolios via trading orders at global financial exchanges. Trading orders are given by the trading decision makers (fund managers, traders etc.) to the execution venues (e.g. brokers, market makers, banks etc.) in order to be fulfilled at the best possible market price. The process of orders matching in the exchanges is governed by several clearance rules, which are different between the global financial exchanges (e.g. NYSE and DAX exchanges clear the agents' orders differently). Clearance rules are focused on securing normal market trading conditions, with smooth and not violent (highly volatile) price actions. Another argument towards the belief that the global trading microstructure affects the market dynamics is that responsible execution venues that govern the majority of financial transactions worldwide haveto fulfill the following requirements. Firstly, the need to follow guidelines and restrictions imposed by state financial authorities. Secondly, they have to be conscientious about their participation rate on the overall daily trading activity, in order not to distort the structural dynamics and to smooth trading operations of traded financial assets. The so called liquidity providers (mostly banks and market making firms) react differently on inflows by other institutional or retail market participants following mostly compliance and risk management rules and guidelines forced by state authorities (e.g. ESMA - European Securities and Markets Authority). All of these decisions and limitations are reflected in the dynamics of the financial assets that we aim to forecast and trade.

Under normal market conditions, the net effect of the aforementioned conditions is that the shifts in the investments between asset classes that are influenced predominantly by supervised institutions must be relatively slow and smooth, as well as interconnected under some invariant internal properties of the entire market itself. Two scenarios would most likely happen if the above assumption did not hold true. Firstly, sudden gaps in the assets' intradaily and daily dynamics would be more frequent - something that is not observed (except from market crashes periods like the 2008 mortgage crisis in US or other similar crashes). Secondly, large financial institutions would (frequently) manipulate markets at will, hence, affecting abruptly the stability and smoothness of the overall trading activity. However, observing such behavior in the long run, would lead to repetitive legal suites imposed by authorities which would lead to frequent insolvencies.

The strong structural limitations imposed by the state authorities to the most influential market participants suggests that there must be a pattern that depicts the effects of several invariant properties of the biggest asset classes dynamics. In addition, market participants follow the rationale of the risk diversification benefits among the several asset classes. This basic investing rationale supports the hypothesis that the intercorrelationdynamics of asset classes are themselves smooth. In this sense, asset classes can be expected to react smoothly to structural forces. Within this framework, we used convolution analysis to find patterns among the representatives of the four biggest financial asset classes that would provide "good" forecasters for indices such as the S&P500.

## 2.2 Dataset

For our analysis we used Bloomberg tickers downloaded via the Bloomberg API from 03 January 2000 until 15 December 2014. We considered the most liquid futures contracts of 42 financial assets traded globally. The assets were selected on the basis of the global GDP ranking across countries worldwide according to the lists published by the International Monetary Fund (IMF). These lists traditionally include the following countries: EU, USA, China, Japan, Germany, UK, France, Brazil, Italy, India, Russia, Canada, Australia, South Korea, Spain and Mexico. Additional asset selection criteria were gleaned from the monthly reports of the World Federation of Exchanges depicting the volume ranking of the global stock exchanges by market capitalization, as well as the volume ranking of the global total value of bonds trading. Following common practices for a representative aggregation of data worldwide, we selected the basic assets for the USA aggregating the America region, the same basic assets for the EU aggregating the European region and the basic assets for China, Japan, Australia and New Zealand for the Asia region. Due to lower trading activity and volume, the emerging markets' dynamics is targeted solely in the FX market (see e.g. investopedia.com, world's most traded currencies by value 2012. 2013; Bank for International Settlements, Triennial Central Bank Survey, 2013).

Our heuristic mining approach to try and efficiently forecast the underlying dynamics of ES1 Index resulted to the following list:

***Equities Markets***: E-Mini S&P 500 (ES1 Index), E-Mini Nasdaq 100 (NQ1 Index), Eurex DJ Euro Stoxx 50 (VG1 Index), Cac40 (CF1 Index), EurexDax (GX1 Index), Ftse 100 (Z1 Index), Ibovespa (BZ1 Index), Swiss Market Index (SM1 Index),

Mexican Market Index (IS1 Index), Australia Market Index (XP1 Index), Nikkei (NK1 Index), Topix (TP1 Index), Hang Seng (HI1 Index)

*Bonds Markets*: 10 Year U.S. T Note (TY1 Comdty), 2 Year U.S. T-Note (TU1 Comdty), Canadian Government 10 Year Note (CN1 Comdty), France Government Bond Future (CF1 Comdty), Eurex Euro Bund (RX1 Comdty), Eurex Euro Schatz (DU1 Comdty), Gilt UK (G1 Comdty), 5 Year T-Note (FV1 Comdty), Japan 10 Year Bond Futures (BJ1 Comdty), Australian 10 Year Bond (XM1 Comdty), Australian 3 Year Bond (YM1 Comdty), Euro Bobl (OE1 Comdty)

*Commodities Markets*: Natural Gas (NG1 Comdty), Gold (GC1 Comdty), Silver (SI1 Comdty), Crude Oil (CL1 Comdty), Corn (ZC1 Comdty)

*FX Markets*: EURUSD Curncy, USDJPY Curncy, EURJPY Curncy, EURCHF Curncy, GBPUSD Curncy, EURGBP Curncy, AUDUSD Curncy, AUDJPY Curncy, NZDUSD Curncy, EURAUD Curncy, USDRUB Curncy, USDCNH Curncy, USDMXN Curncy, USDINR Curncy

The dataset records are daily futures closing prices of the futures contracts. Due to the fact that most of these assets have different trading time zones and closing hours, the prices were preprocessed in a way that the final dates used for the analysis corresponded on closing prices for all the dataset's assets. A second transformation was required in order to address the problem of non-comparable spot prices. To normalize the data we used the following cumulative return formula to rebase every asset on the unit base and to compare their overall dynamics with price returns terms:

$$P_{new} = 1 + \sum_{i=1}^{n} dlog(P_{old}), \qquad (1)$$

where $P_{new}$ are the new rebased prices, $P_{old}$ are the raw time series data downloaded from Bloomberg API, and dlog is the operator that provides the returns of the raw assets prices.

**2.3 Data Analysis**

The first step involves the generation of the whole set of convolutions of all future contracts of the financial assets with the ES1 Index - future contract of S&P500 using the discrete convolution equation:

$$(f * g)[n] \stackrel{def}{=} \sum_{m=-\infty}^{\infty} A_t^i[n-m]ES_t[m], i = 1,..,42 \qquad (2)$$

where $A_t^i$ denotes the i-th (of the total 42) future contract rebased price time series according to equation (1) and denotes the rebased price time series of the ES1 Index. For our analysis we tested the forecasting efficiency of the coarse-grained variable

$$W_t = \sum_{i=1}^{42} E[d(P_i)], \qquad (3)$$

where $E[d(P_i)]$ is the mean value of the rebased prices for each of the 42 assets. We show that this coarse-grained variable can be used to trade the S&P500 future contract with better results than the "Buy and Hold" benchmark strategy, offering therefore a relatively simpler approach compared to other quantitative trading methods. Using $W_t$, the profit over time is calculated by:

$$\Pi_t = \int_0^t W_t * d(ES_t), \tag{4}$$

Equation (4) above expresses the convolution between $W_t$ and $d(ES_t)$, i.e. the traded asset's returns.

The prevailing assets selection is done using the best Sharpe Ratio as the basic criterion for the most influential assets extraction. The mean value operator in the $W_t$, variable is estimated using simple moving averages on the normalized prices, approximating a trading rationale used by many technical traders as well as quantitative analysts practice trying to spot trending or mean reverting periods in the price dynamics.

We use three different moving averages approaches targeting a specific period pattern cyclicality. In particular we used Exponential Moving Averages (EMA) with three separate time lags (trading days) as inputs: 3, 25 and 500 days. Using these three lags we aimed at capturing different convolution dynamics patterns of the $ES_t$ relative to the whole other assets' universe targeting daily, monthly and 2-year trading activity, respectively.

Following the above rationale, we rewrite the variable $W_t$ as:

$$W_t = \sum_{i=1}^{42} sign(EMA_L[d(P_i)]), \tag{5}$$

with L being the lag period (L= 3, 25, 500 days). The *sign* function is used to filter noise effects of skewness and kurtosis emerged by the assets' probability distribution, thus focusing only on the positive or negative jumps of the price movement in a binary sense (1 for positive, -1 for negative, 0 for neutral). The forecasting power of this simple approach is enforced by using trading simulators the results of which are compared to the "Buy and Hold" benchmark strategy.

The above analysis leads to a wide spectrum of pairs under the convolution rationale between *ESt* and the remaining assets. The overall procedure seems to be a common "cherry picking" approach - but still the total process is based only on representative assets dynamics that are used to forecast the future contract under study. Thus, our procedure considers the dynamics of the "entire" market to forecastone of its subsets. The hypothesis is that fundamental structure changes will trigger changes in the assets themselves as well as in their interconnected dynamics. Due to authorities limitations in the trading activity among majorcounterparties as well as the correlations between the several basic asset classes riskpremia, a realistic assumption and belief is that strong and violent shifts in investorpreferences will be rare to occur in relatively short periods. If this structural connection between the assets is made, then we "believe" that a transition will be slow enough so that it can be captured by the same forecast indices.

**2.4 Trading Simulations**

Trading simulations are based on the resulting profit and loss (PnL) of a portfolio with an initial investment of 1M Euros permitted to be invested only in the ES1 Index. The trading signal is shifted by one trading day, i.e. $W_{t+1} = S(W_t)$, where $S$ is the shift operator. A simple risk management method is used to include the effects of common capital preservation and risk discipline used by most of the trading practitioners in the market. The following steps describe the simple procedure that implements the trading strategy:

1. Calculate $W_t$ from each asset in the dataset including the ES1 Index.
2. Multiply each of these signals with the ES1 Index time series and apply the cumulative sum operator. Denote the result as $X_i, i = 1, \ldots, 42$
3. Apply the "Best Sharpe Ratio" approach to select a subset (or all) of the $X_i s$. Formulate an equally weighted signal of the selected forecasters. This is the final trading strategy on S&P500 futures contract ES1 Index.
4. Risk Calibration. Calculate the rolling annualized volatility of the strategy's returns at each time step divide the trading signal (weight) with it, targeting a 1% volatility in the strategy's final returns. The annualized volatility is defined as

$$AnnualizedVolatility = sqrt(252) * std(Returns), \qquad (6)$$

for the 252 trading days.

Here, the most influential assets in the ES1 Index forecasting were selected based on the Best Sharpe Ratio. In step (3) above we are able to set a predefined number of basic forecasters of the ES1 Index based on their efficiency to produce consistent trading signals since inception. More specifically, we extract the assets (*N* of the whole dataset) that produce the best "absolute" Sharpe ratio calculated on the $X_i$ time series, where "absolute" means that we are able to reverse the signal's sign if the $X_i$ for this particular asset produces a largely negative Sharpe ratio. That way, we distinguish between trend following and mean reversion; the purely positive Sharpe ratios indicate a good trend indicator for ES1 Index while negative Sharpe ratios provide evidence for mean reverting leaders.

## 3. RESULTS AND DISCUSSION

The performance of the trading strategy for the ES1 Index - future contract (shown in Figure 1),was tested using four different sets of forecasters (N= 10, 20, 30, 42) and three time lags (L= 3, 25, 500).

Figure 2 shows the cumulative returns performance,$\Pi_t$,of the ES1 Index trading strategy when selecting 10 optimal (with respect to the Best Sharpe Ratio) assets as forecasters, using three lags (3 days, 25 days and 500 days). In the period of high volatility due to well-known financial crises (e.g. in 2008 with the Lehman Brothers and in 2011, with discussions on an impending Grexit), the strategy internally shifted close to zero risk allocation (Figure 3 a,b,c) for all three lags, while it maintained relatively small drawdowns in the overall cumulative performance (Figure 2). Furthermore, the proposed change of weights (Figure 3 a,b,c) experience relatively volatile changes between investment periods t and t+1 for lags 3 days and 25 days, while is "smoother" for the lag of 500 days as expected. As it is shown in Figure 3, the cumulative returns performance of the portfolio corresponding to the lag of 500 days is close to the portfolio with the lag of 25 days (both being relatively far from the one with the lag of 3 days).

Figure 4 depicts the cumulative returns, when choosing 20 optimal (with respect to the Best Sharpe Ratio) forecasters for the ES1 Index. As it is shown, now the Cumulative Returns performance of the portfolio corresponding to the lag of 500 days is clearly far from the portfolios with the lags of 3 and 25 days. The corresponding weights for the 20 optimal forecasters portfolio is shown in Figure 5 a,b,c for the three lags.

Figure 6 shows the cumulative returns for 30 forecasters. All three portfolios' dynamics are close to each other in terms of cumulative returns performance, while they exhibit differentfrequencies of weight allocation from day to day investment interval (Figure 7 a,b,c): the dynamics with the lag of 500 days experiences a "smoother" weight allocation (Figure 7 c).

Figure 8 shows the cumulative returns when we used the entire proposed dataset in order to forecast the ES1 Index. The results suggest that the cumulative performances for all three portfolios (corresponding to the three different lags) are stable, robust, beating the ES1 Index's performance. All three portfolios stay close to each other in terms of the cumulative returns, while the weight allocation frequency (even for the case of the lag of 500 days) is higher when compared to the other sets (10, 20 and 30 forecasters). Figure 9 a,b,c shows the corresponding weights for this "complete" portfolio.

Finally, Table 1 summarizes the Sharpe ratios as obtained by the proposed methodology under the different scenarios with respect to different EMA lags as well as to different number of forecasters. The Buy and Hold strategy's Sharpe ratio of the ES1 Index defined on the raw returns of the index itself is 0.10 for the period under study. Hence, the proposed methodology outperforms consistently the "Buy and Hold" benchmark. Thus, these results depict that combinations of long term dynamics/ basket of forecasters produce a robust and consistent way of generating significant returns when trading the ES1 Index. The values of the Sharpe ratios can also help indicate the best combination using fundamental analysis on the optimal forecasters.

Acknowledgements: P.P. would like to acknowledge fruitful discussions with NikolaosParliaris - The Olayan Group, Athens, Greece

**Table 1.** The Sharpe Ratios as computed using the proposed trading approach with respect to different EMA lags (3, 5, 500) as well as different number of forecasters (10, 20, 30, 42).

| Sharpe Ratios | | | |
|---|---|---|---|
| # Forecasters | EMA(3) | EMA(25) | EMA(500) |
| 10 | 0.75 | 0.55 | 0.82 |
| 20 | 0.43 | 0.45 | 0.80 |
| 30 | 0.48 | 0.58 | 0.60 |
| 42 | 0.67 | 0.9 | 0.70 |

**Figures**

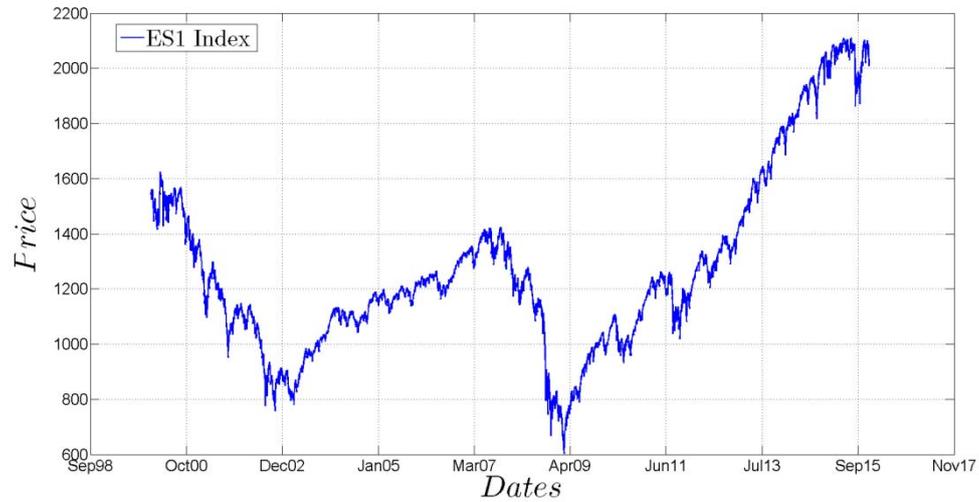

**Figure 1.** The ES1 Index spot price in the time period 03 January 2000 -15 December 2014.

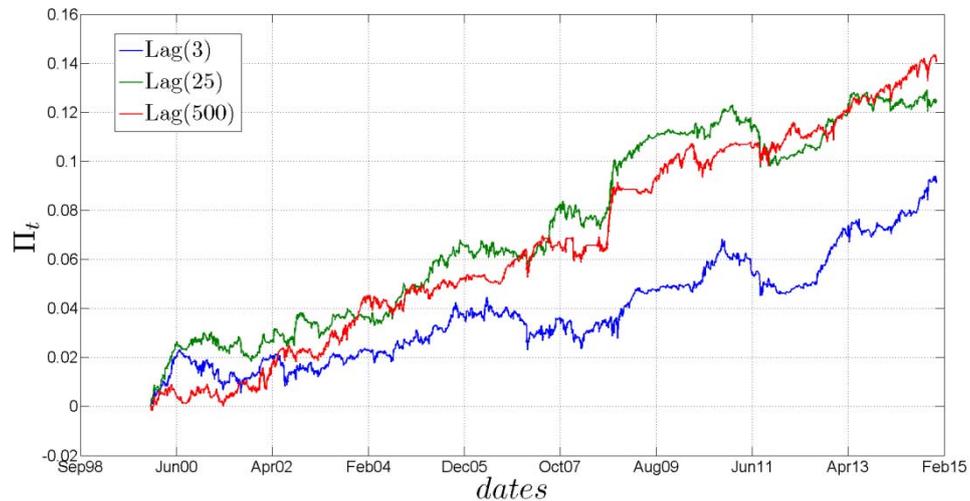

**Figure 2.** The Cumulative Returns performance, $\Pi_t$, of the strategy trading the ES1 Index using 10 optimal assets as forecasters for the three different time lags (3, 25, 500 days).

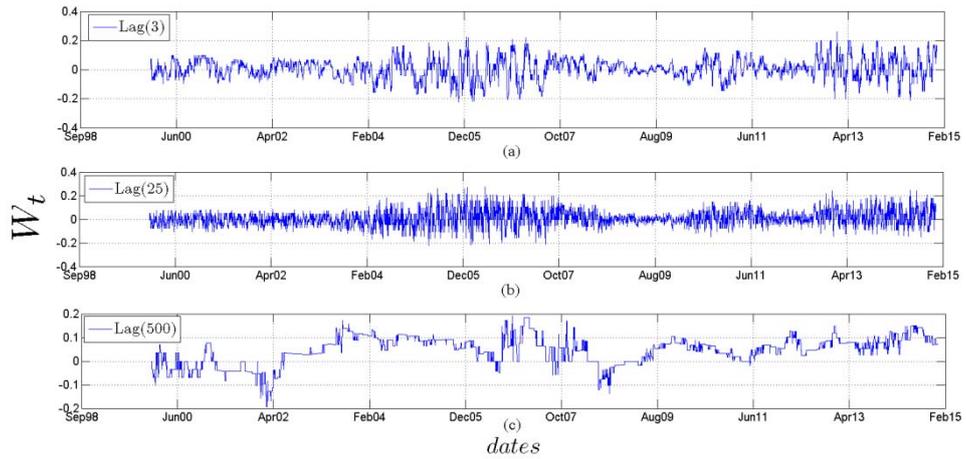

**Figure 3.** The weights, W$_t$ of the strategy trading the ES1 Index using 10 optimal forecasters for a time lag of (a) 3 days, (b) 25 days, (c) 500 days.

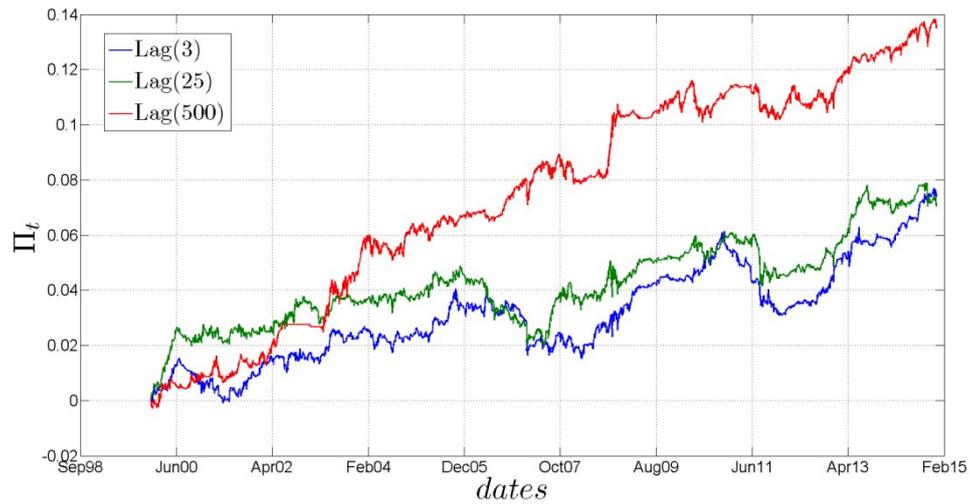

**Figure 4.** The Cumulative Returns performance, Π$_t$, of the strategy trading the ES1 Index using 20 optimal assets as forecasters for the three different time lags (3, 25, 500 days).

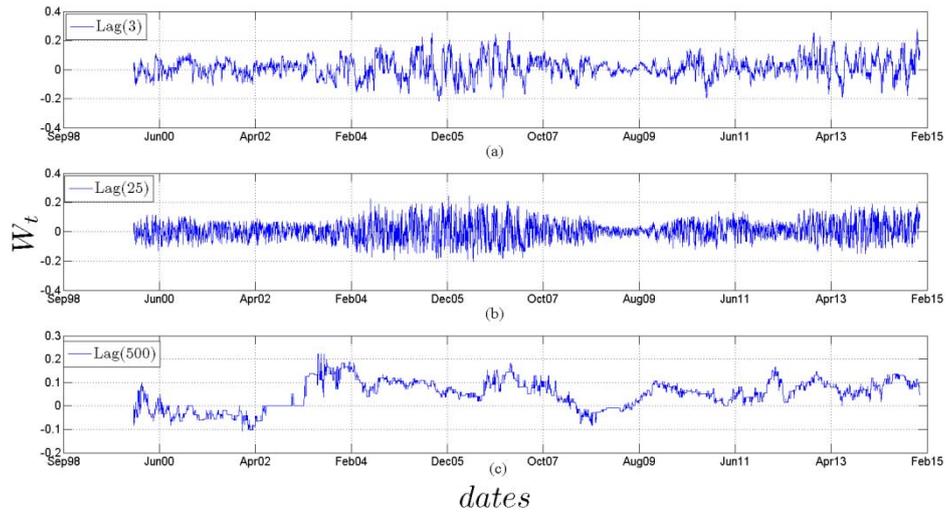

**Figure 5.** The weights, $W_t$, of the strategy trading the ES1 Index using 20 optimal forecasters for lag of (a) 3 days, (b) 25 days, (c) 500 days.

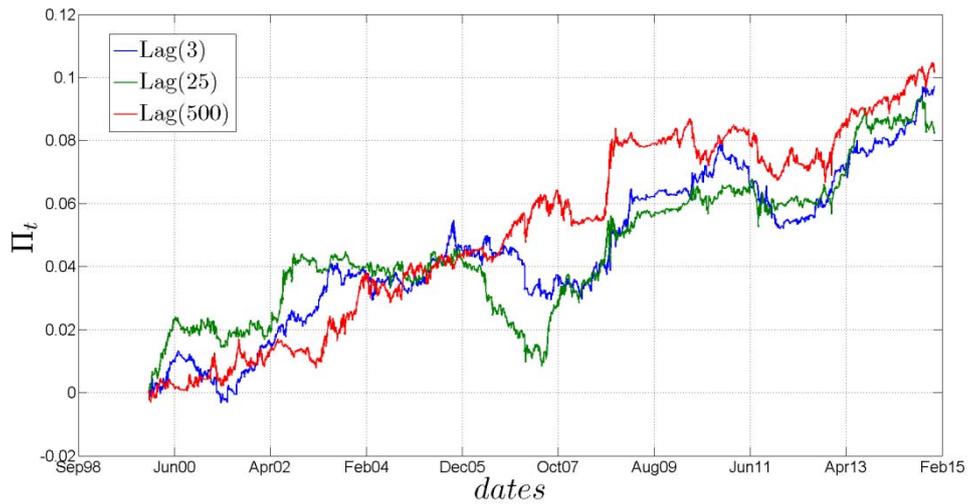

**Figure 6.** The Cumulative Returns performance, $\Pi_t$,, of the strategy trading the ES1 Index using 30 optimal assets as forecasters for the three time lags (3, 25, 500 days).

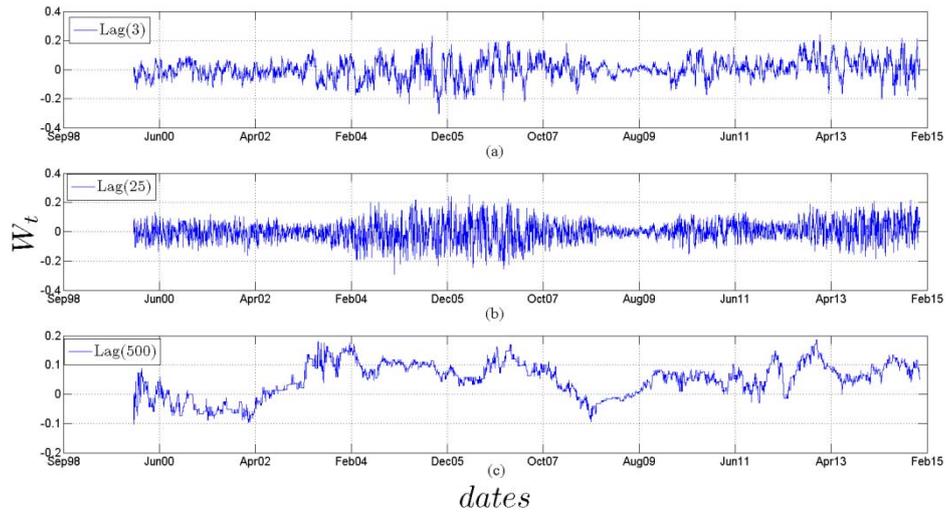

**Figure 7.** The weights, $W_t$, of the strategy trading the ES1 Index using 30 optimal forecasters for a time lag of (a) 3 days, (b) 25 days, (c) 500 days.

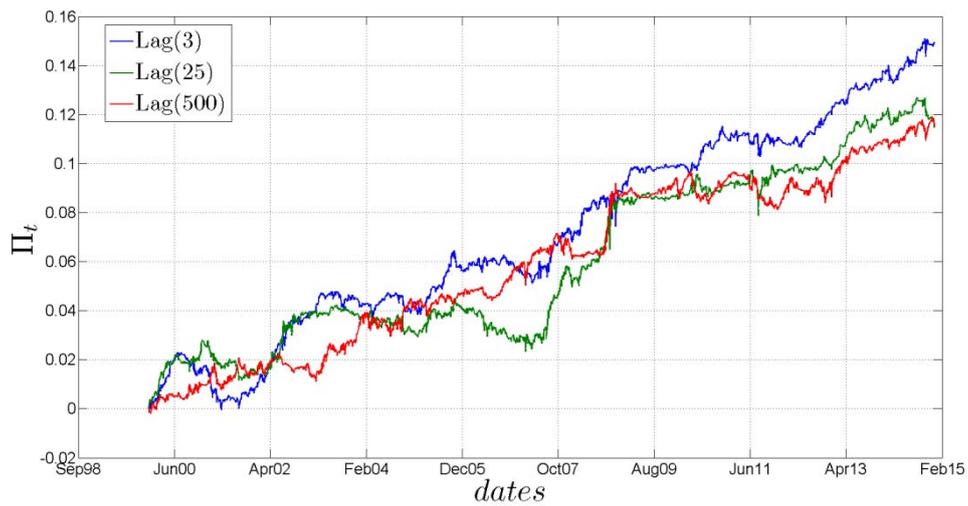

**Figure 8.** The Cumulative Returns performance, $\Pi_t$, of the strategy trading the ES1 Index using the entire assets universe as forecasters for the three different time lags (3, 25, 500 days).

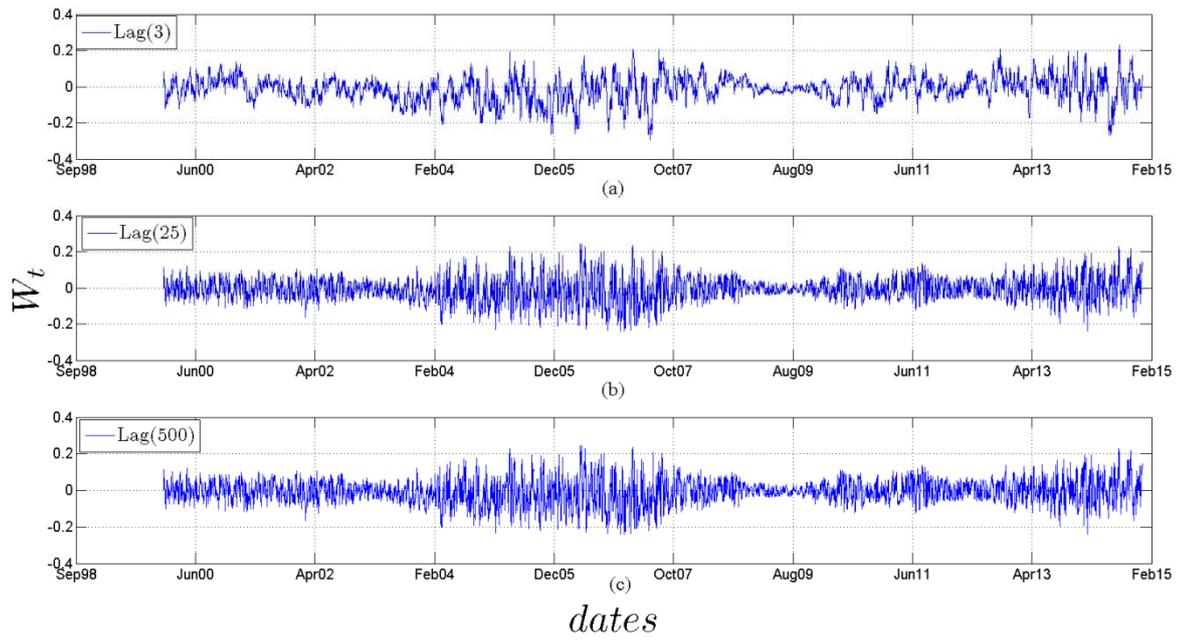

**Figure 9.** The weights, $W_t$, of the strategy trading the ES1 Index using the entire assets universe as optimal forecasters for a time lag of (a) 3 days, (b) 25 days, (c) 500 days.